# Authentication and Billing Scheme for The Electric Vehicles: EVABS[1]

Ömer AYDIN[2]

**Abstract**

*The need for different energy sources has increased due to the decrease in the amount and the harm caused to the environment by its usage. Today, fossil fuels used as an energy source in land, sea or air vehicles are rapidly being replaced by different energy sources. The number and types of vehicles using energy sources other than fossil fuels are also increasing. Electricity stands out among the energy sources used. The possibility of generating electricity that is renewable, compatible with nature and at a lower cost provides a great advantage. For all these reasons, the use of electric vehicles is increasing day by day. Various solutions continue to be developed for the charging systems and post-charge billing processes of these vehicles. As a result of these solutions, the standards have not yet been fully formed. In this study, an authentication and billing scheme is proposed for charging and post-charging billing processes of electric land vehicles keeping security and privacy in the foreground. This scheme is named EVABS, which derives from the phrase "Electric Vehicle Authentication and Billing Scheme". An authentication and billing scheme is proposed where data communication is encrypted, payment transactions are handled securely and parties can authenticate over wired or wireless. The security of the proposed scheme has been examined theoretically and it has been determined that it is secure against known attacks.*

***Keywords:*** *electric land vehicle, authentication scheme, encryption, security, payment system, charging system.*

---

[1] Bu çalışma, 14-15 Ekim 2020 tarihlerinde 4. Uluslararası Ekonomi, Finans ve Enerji Kongresi'nde sunulan bildirinin genişletilmiş halidir.
[2] Dr. Öğr. Üyesi, Manisa Celal Bayar Üniversitesi, Mühendislik Fakültesi, Elektrik Elektronik Mühendisliği Bölümü, Bilgisayar Bilimleri Anabilim Dalı, omer.aydin@cbu.edu.tr

# 1. INTRODUCTION

Since the use of electricity by people, this energy has been the main source of many technological developments. Televisions, phones and computers are some examples of this development. The energy required for the operation of many technological inventions and devices is electricity. In this development process, the use of electricity among other energy sources has increased day by day. Since the development of the first steam-powered land vehicle (Reitze Jr, 1977) by Nicholas Cugnot in 1769, serious changes have occurred in land, sea, air vehicles and spacecraft. Due to the adverse effects of fossil fuels used in internal combustion engines such as environmental pollution, their limited amount in the world and the prediction that they will be depleted in the near future, the need to focus on different energy sources has arisen.

Fossil fuel reserves are decreasing rapidly and especially oil and natural gas reserves are approaching critical levels (Öztornacı, 2019). The distribution of energy demand by years in terms of distribution by resources is given in Table 1.

**Table 1.** Distribution of primary energy demand by resources (million tons of oil equivalent)

|             | 1990   | 2010   | 2015   | 2020   | 2030   | 2035   |
|------------:|--------|--------|--------|--------|--------|--------|
| Oil         | 3.2300 | 4.113  | 4.352  | 4.457  | 4.578  | 4.656  |
| Coal        | 2.231  | 3.474  | 3.945  | 4.082  | 4.180  | 4.218  |
| Natural gas | 1.668  | 2.740  | 2.993  | 3.266  | 3.820  | 4.106  |
| Biomass     | 903    | 1.277  | 1.408  | 1.532  | 1.755  | 1.881  |
| Nuclear     | 526    | 719    | 751    | 898    | 1.073  | 1.138  |
| Hydraulic   | 184    | 295    | 340    | 388    | 458    | 488    |
| Other       | 36     | 112    | 200    | 299    | 554    | 710    |
| Total       | 8.779  | 12.730 | 13.989 | 14.922 | 16.417 | 17.197 |

**Source:** Kalkınma Bakanlığı Özel İhtisas Komisyonu Raporu, 2014; Aydın, 2020

When we consider all this information, we see that along with electrical energy, hydrogen, biological fuels, and nuclear reactions energy sources are used. Considering its ease of use, accessibility and various advantages in terms of production, electrical energy stands out among the energy sources used by the vehicles used in our daily life.

Although the use of electric vehicles began in the late 19th and early 20th centuries, electric vehicles began to take place more in today's world by developing and increasing in prevalence after 1980 and especially in the early 2000s. Electric bicycles, cars, trains, buses and planes are now becoming widespread seriously. The necessity of filling the batteries of electric bicycles and cars in the individual vehicle class within these vehicles and the difficulties brought by this requirement appear as issues that need to be solved. There is a need for a safe and standard charging system for charging vehicles, especially in countries where individual parking facilities are very low. Charging vehicles parked on the street and using information technologies during this process, charging, device authentication and billing. Information systems are needed to work on these issues. There is no standard established for this new situation. Therefore, it is necessary to carry out studies that can direct the establishment of the standard.

The aim of this study is to provide a solution for charging electric vehicle batteries while parked on the streets. With the solution offered, transactions will be defined for payment of the billing process, mutual authentication of the devices, secure communication, termination of the billing process and invoicing. These definitions and solutions, it is aimed to contribute to the creation of a standard for battery charging processes of electric vehicles.

# 2. LITERATURE REVIEW

Many key agreement and authentication schemes have been proposed and examined for vehicles using IoT (Lo et al., 2015; Kumari et al., 2016; Chin et al., 2017; Liu, et al., 2017; Mohit et al., 2017; Guo et al., 2017; Zhou et al., 2017; Shen et al., 2017; Wu et al.,2019). Although these studies have positive results in terms of privacy and efficiency, they are vulnerable to some attacks. Considering the resource constraints of the IoT, key exchange or authentication schemes that are resistant to existing attacks should be proposed.

On the other hand, charging electric vehicles is a serious and important issue for smart grid systems. Smart grids should be efficient, secure, stable and sustainable. Many studies have been done in the literature on charging electric vehicles in smart grids (Kim et al., 2010; Lu et al., 2012; Gan et al., 2012; Xu et al., 2016; Tian et al., 2016; Tang & Zhang, 2016). Some security and privacy problems may be encountered in these studies. Some studies have suggested the use of blockchain to solve these problems (Nakamoto, 2008; Surhone et al.,2010; Poon & Dryja, 2016; Karaarslan & Akbaş, 2017; Sarıtekin et al., 2018; Huang et al., 2018; Karaarslan & Adiguzel, 2018; Hyperledger Foundation, 2022; NIST Series Pubs., 2022).

Vaidya et al. have proposed a comprehensive and decentralized authentication mechanism. This mechanism aimed to eliminate the flaws and deficiencies in central authentication systems and to provide strong authentication in a multi-server environment. The study includes functional requirements, performance and security analysis. It has been claimed that with the proposed mechanism, the authentication delay decreases compared to the existing schemes and a more robust and efficient solution is proposed (Vaidya et al., 2011).

As it is known, electric vehicle charging becomes a part of our daily life. It is usually done in residential or public areas. Roberts et al. examined electric vehicles using authentication protocols during charging. They proposed a new protocol that uses key exchange without trusting certificates in their study. They have implemented this protocol using Bluetooth and Wi-Fi and claim to provide the necessary communication framework (Roberts et al., 2017).

Li et al. proposed an authentication protocol named Portunes+. This protocol is implemented using Raspberry Pi2. They made various time measurements for the authentication process. In their security analysis, they obtained numerical results showing that external attacks have decreased. They claimed that the protocol they proposed was both reliable and computationally efficient (Li et al., 2016).

Central charging servers are vulnerable to attacks such as Distributed Denial of Service and Insider Threats by attackers who have privileged access. As it is mentioned earlier, blockchain-based systems have been proposed to fix these problems. In some studies, key security cannot be guaranteed. Kim et al. proposed a blockchain-based system. Key security, mutual secure authentication, confidentiality and an efficient structure are promised in this system. Secure mutual authentication has been proven with BAN logic. Resistance to replay and man-in-the-middle attacks has been verified with AVISPA. They have added comparisons for computing and communication costs. They claimed that this system was suitable for lightweight devices (Kim et al., 2019).

## 3. SECURE PAYMENT AND CHARGING SYSTEM

In countries where individual parking areas are not sufficient, the ability to charge people's vehicles while the devices are parked on the streets, shopping malls and other parking areas will provide many advantages. Especially when it is considered that many people leave their vehicles until they return home after work in the evening and park on the streets in the morning or on the weekends, it is of great importance to be able to charge their electric vehicles safely during this period. Vehicles in apartments without individual car parks are open to interference from outside during charging. In this process, disconnecting the charge, connecting to another device without interrupting the charge with intervention and billing to the wrong person, etc. There are many risks. For all these reasons, the authentication scheme and invoicing system, details of which are shared below sections, have been designed.

### 3.1. Current charging and payment technology

According to the International Energy Agency, electric vehicles account for 2.6% of global automobile sales, according to 2019 data. In terms of sales, there was a 6% growth compared to 2018 (McBain & Bibrah, 2021). Along with these developments, electric vehicle charging systems and infrastructure have been developing and expanding in the last 10 years. For North America, there are at least 10 EV charge point operator(CPO)s (Cision PR Newswire, 2020). If we look at the situation in the United States, more than 60,000 individual charging outlets and more than 20,000 charging locations serve (Vehicle Technologies Office, 2019). In line with all developments, the EV charging industry is expected to have a value of over $100 billion by 2027(Cision PR Newswire, 2020).

Electric vehicle charging systems are classified as Level 1, 2 or 3 depending on current type, charging speed and connectors. Level 1 uses a conventional household electrical outlet that provides up to 15 amps (A) of AC current at 110 volts (V) and a maximum speed of 1.65 kilowatt hours (kWh) (Firewire Tech, 2020). Level 2 EV chargers also supply AC household current, but use a two-phase connector and upgraded cable, allowing EVs to charge up to 30A–50A at 240V (homes) and 80A in commercial locations (Charge Point, 2021). Level 3 EV charging systems, also known as DC fast charging, are the closest option to refueling a petrol vehicle. DC fast chargers have been significantly improved and currently offer charging rates between 50kWh–350kWh (Charron, 2019), with higher rates expected in the future. These chargers are more complex and convert electricity from AC to DC, bypassing the vehicle's built-in converter, which usually limits Level 2 charging speeds.

The whitepaper titled "Electric Vehicle Charging Open Payment Framework with ISO 15118" introduced a framework where an electric vehicle (EV) driver can charge the vehicle at a charging station located at any location and pay without a membership or account. Detailed explanations and details of a safe charging and payment system are shared in the relevant document. This white paper proposes the use of the ISO 151181 Vehicle-to-Grid communication standard along with open payment technology. With Plug&Charge (PnC) ISO 15118, the system comes into play and safe EV charging can be implemented in a simple way. This document presents an approach to safe EV charging with the support of open payment technology(S.T.A., 2021).

Vaidya and Mouftah says taht ISO/IEC 15118 standard defines certificate-based authentication and authorization mechanism for EV charging, but since these techniques are single-mode and single-path, they have low security and may be vulnerable to various malicious attacks. Multi-mode and Multi-Pass Authentication mechanisms have been proposed for EV networks. They claimed that substitution attacks, Man-in-the-middle attacks etc. can be reduced by using such mechanisms (Vaidya & Mouftah, 2020).

Wang et al. proposed a lightweight authentication framework based on federated learning (FL) with a premium penalty mechanism for EV infrastructures considering both the efficiency of energy management and the potential risks of federated learning (FL) for EV infrastructures. They claim that the security analysis and performance evaluation proves that their proposed framework can create an accurate electricity demand forecasting framework for EV infrastructures to defend against multiple federated learning (FL) attacks (Wang et al., 2021).

Babu et al. proposed a protocol package EV-Auth for entities communicating in a dynamic charging system capable of mutual authentication and session key calculation. To gain practical capabilities, EV-Auth is built to rely on lightweight encryption primitives. It was emphasized that it includes a feature known as uninterrupted electric vehicle delivery. The security of the proposed protocol was evaluated using the random oracle model and the scyther tool(Babu et al. 2022).

### 3.2 Proposed payment and charging system

The parking of vehicles on the streets and the representative image of the proposed street charging and billing system are shown in Figure 1. It is assumed that the vehicles will be placed in the pockets marked on the street in the figure. There are connection points right next to the pockets and on the pavement where the cables to be used for charging are located. When the cables here are pulled, they can be extended by unwinding from their windings by means of the mechanism inside. When it is released, it is collected. Most of these systems are used in the electrical cables of vacuum cleaners. The electrical cables are designed to be resistant to external interference, to eliminate the risks such as leakage and short circuit, and to have outer coatings. These electrical cables are connected to terminals located opposite each other and located at the endpoints of each street. These terminals are connected to the main servers with secure infrastructure and are not open to intervention. It is assumed that the communication is encrypted and the information transmitted is secure and cannot be changed.

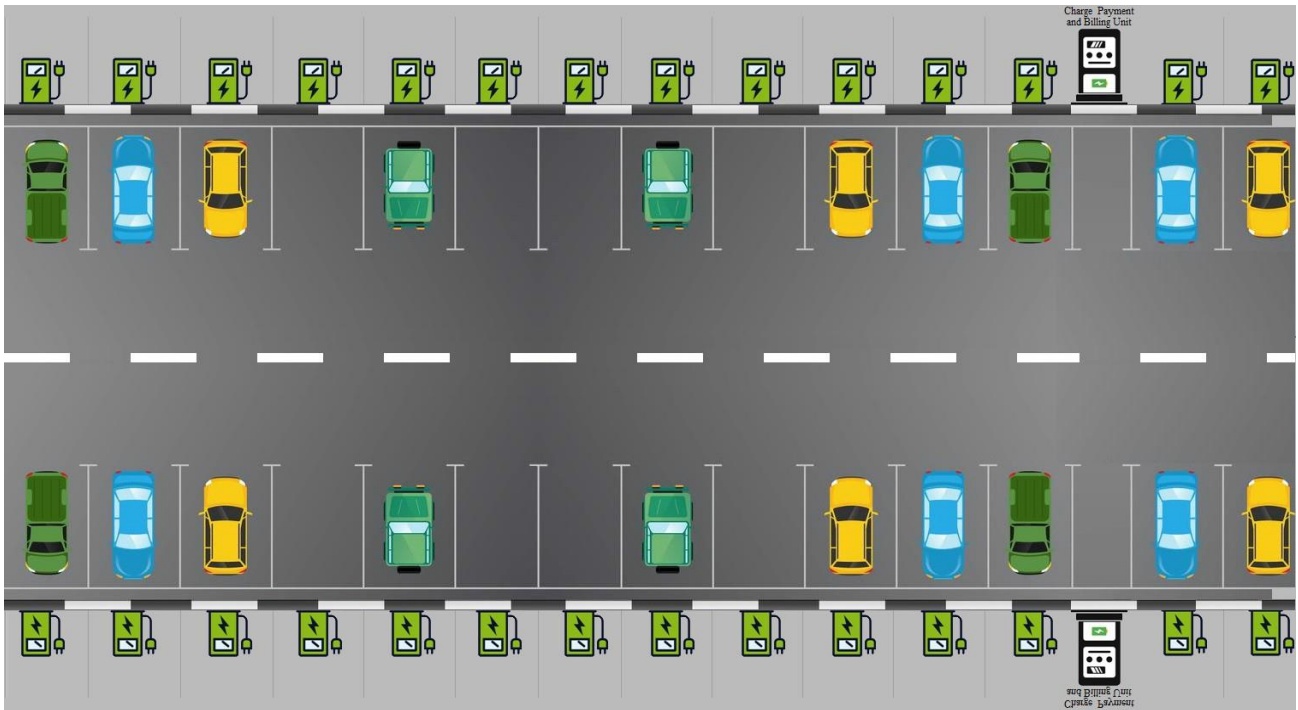

**Figure 1.** View of the charging and payment system components on the street

When the vehicle approaches the parking pocket, the charging cable will be pulled out and attached to the vehicle. Once installed, the authentication process will be provided between the terminal and the vehicle, regardless of whether it is wired or wireless. During or before the authentication process, the owner of the vehicle enters the amount he wants to load through the terminal and then pays the payment by credit card, cash, etc. will be able to. In addition, it will have the opportunity to load the balance into its predefined account and recharge the vehicle from the current balance from its mobile device or any compatible device connected to the internet. The relevant terminals are designed to allow all these payment transactions. At the same time, the terminal will cut the charge at the end of the charge and the server will send the relevant billing information to the owner of the vehicle by text message, e-mail, etc. will transmit according to preference. In addition to these, if desired, it can be printed on the street with the possibility of printing the plug on the terminal.

Performing authentication can be considered as a mutual handshake between the vehicle and the server. In parallel with its use in daily life, very important functions such as agreement, mutual recognition and trust are fulfilled with the authentication process, which is described as a handshake. In addition, by creating a framework such as the stages of this process between the two devices, which parties and transactions it has, this turns into a scheme. An authentication scheme given in Figure 2 has been proposed for this process, and its implementation is supposed to protect against known attacks.

Figure 2 shows the communication between the Server, the Charging Station and the Electric vehicle. The information transferred between them is shown with arrows for the devices to verify each other and for the billing process. The authentication and billing scheme shows the transactions between the server and the terminal and the electric vehicle. There is an insecure connection between the Terminal and the Vehicle. First, the vehicle encrypts a random number and its own ID and sends the encrypted text with the random number to the terminal. The terminal decrypts it and sends it to the server. The server is checking database records with relevant information. If the vehicle is not in the list, the authentication failed message is transmitted to the vehicle via the terminal. If the authentication is successful, the charge process start message is transmitted to the terminal. In this way, the charging process begins. If the charging process finishes or is interrupted, this is detected and time information is transmitted to the server by the terminal. In this way, the server performs the billing process to the relevant vehicle.

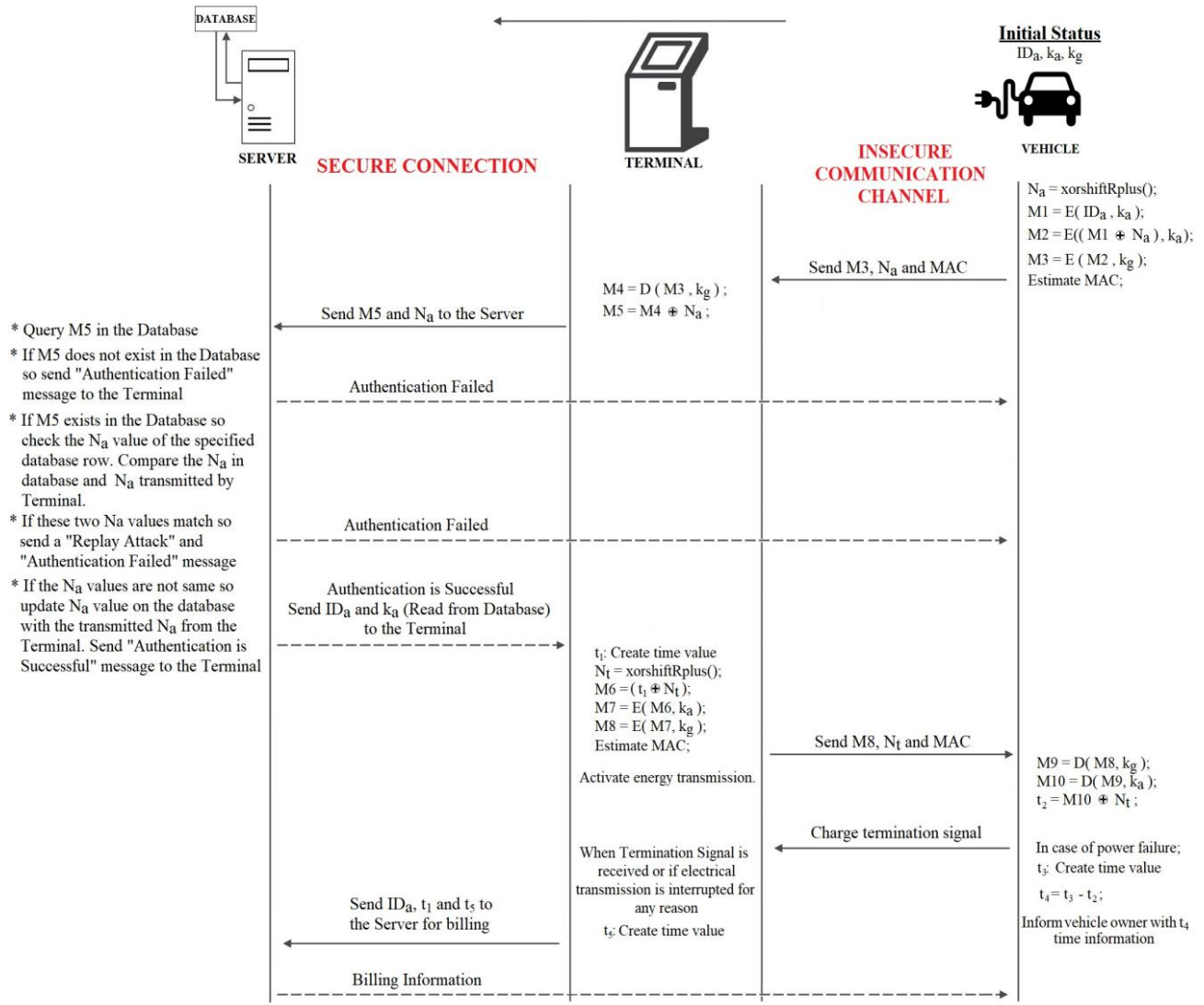

**Figure 2.** Authentication and billing scheme

Before examining the authentication scheme steps in detail, let's share the information of the parties of the system and general information about the status of the system. On the vehicle, there are IDa, ka and kg recorded by creating records on the server beforehand. The tool and terminal xorshiftRplus (Cabuk et al., 2017) can generate random numbers using the pseudorandom number generator (PRNG). For systems with lower resources, xorshiftULplus (Aydın & Kösemen, 2020) PRNG can be recommended. However, since it will generally have sufficient resources in terms of vehicle and terminal hardware and connection source, xorshiftRplus, which provides higher security, was used in this study. It has a value of kg to use as well as terminal encryption. These symbols and notations are given in Table 2 with their explanations.

**Table 2.** Signs and symbols used

| | |
|---|---|
| $k_g$ | Secret group key for group members |
| $k_a$ | Vehicle secret key |
| $ID_a$ | Vehicle identification code |
| $N_a$ | Random number generated on the vehicle side |
| $N_t$ | Random number generated on the terminal side |
| MAC | Message authentication code |
| $\oplus$ | Exclusive or operator |
| $E(X, k)$ | Encrypting X with the secret key k using AES |
| $D(E, k)$ | Decryption of X with the secret key k using AES |
| t | Time value |
| M | Calculated message values |

The attacker is assumed to have the following capabilities, access, information, and resources (Aydin et al., 2020):

- The attacker can listen to all messages between the vehicle and the terminal.
- Attacker can block the data transmitted on the communication channel.
- Attacker can send a message to the other party as if it were a vehicle or a terminal.
- The attacker does not have access to hidden parameters but can access all functions or processes. Pseudorandom number generator, encryption and XOR.
- The attacker can read, create, modify, and delete all messages transmitted and return those messages back to the communication channel.

The authentication scheme steps that we assume to be secure under the assumption that the attacker has these capabilities are as follows.

**Step 1 (Vehicle):**

- Generate the $N_a$ random number (RN) using xorshiftRplus PRNG.
- Vehicle identification code $ID_a$ is encrypted using AES algorithm with $k_a$ key and M1 is created.
- M1 and $N_a$ are subjected to logical Exclusive OR operation and M2 is formed.
- M2 is encrypted with AES using $k_g$ group secret key and M3 is generated.
- MAC is calculated.
- M3, MAC and $N_a$ are sent to the terminal.

**Step 2 (Terminal):**

- M3, MAC and $N_a$ transmitted from the vehicle are received.
- M3 value is decrypted with AES using $k_g$ group secret key and M4 is obtained.
- M4 and $N_a$ are subjected to logical Exclusive OR operation and M5 is formed.
- M5 and $N_a$ are transmitted to the server over a secure line.

**Step 3 (Server):**

- M5 and $N_a$ transmitted by the terminal over the secure line are received.
- Search for M5 in the database.
- If the record is not found, the authentication failed message is returned to the Terminal.

- If the M5 value is found, the $N_a$ value of the relevant database record is checked.
- If the received $N_a$ value and the $N_a$ value on the database are the same, it is evaluated as a replay attack and the authentication operation failed message is returned to the terminal.
- If the $N_a$ value received from the terminal and the $N_a$ value on the database are not the same, the value in the database is updated with the new value transmitted from the terminal, and the $ID_a$ and $k_a$ information in the relevant database record is sent to the terminal over the secure line.

**Step 4 (Terminal):**

- Terminal gets $ID_a$ and $k_a$ values from the server.
- It takes the time value of $t_1$ with the help of the built-in clock.
- Generate the random number $N_t$ (RN) using xorshiftRplus PRNG.
- $t_1$ and $N_t$ is subjected to logical Exclusive OR operation and M6 is formed.
- M6 is encrypted using the AES algorithm with the $k_a$ key and M7 is created.
- M7 is encrypted using the AES algorithm with the $k_g$ key and M8 is created.
- MAC is calculated.
- Energy flow is activated.
- M8, MAC and $N_t$ are sent to the vehicle.

**Step 5 (Vehicle):**

- M8, MAC and $N_t$ sent from the terminal are received.
- The value of M8 is decrypted with AES using the $k_g$ group secret key and M9 is obtained.
- M9 value is decrypted with AES by using $k_a$ vehicle secret key and M10 is obtained.
- M10 and $N_t$ are subjected to logical Exclusive OR operation and $t_2$ is obtained.
- Waits for the completion of the charging process.

In case of power cut for any reason (battery charging, reaching the filling balance, removing the cable manually, etc.)

**Step 6 (Vehicle):**

- The vehicle creates $t_3$ time value with the clock on itself.
- Calculates the time value $t_4$ by subtracting the time value $t_2$ from the time value $t_3$.
- It shows the $t_4$ time value to the owner of the vehicle to inform him/her and saves it to the relevant place in the vehicle.

**Step 7 (Terminal):**

- Generate time information $t_5$ and send start ($t_1$) and end time ($t_5$) information to the server along with the associated vehicle information ($ID_a$).

**Step 8 (Server):**

- Perform the invoicing process with the vehicle information and time information coming from the terminal and deliver the necessary information to the vehicle owner through preferred channels.

## 4. SECURITY ANALYSIS OF THE EVABS

The security of the proposed authentication scheme will be analysed in this section. The security of EVABS against well-known attacks will be evaluated theoretically. Before making this assessment, we will list below the capabilities of the attacker (Dolev & Yao, 1983):

- The attacker can listen to the communication between the terminal and the vehicle and see all the messages.
- The attacker can block the messages between the parties.
- The attacker can replace one of the parties and send a message to the other.
- The attacker knows the encryption algorithms, operations and functions.
- The attacker does not have access to the hidden parameters.
- The attacker can delete, modify or create all transmitted messages and send them again to the communication environment.

EVABS has been examined against to traceability, physical disclosure, replay, impersonation, denial of service, desynchronization, man in the middle, eavesdropping and cloning attacks.

### 4.1. Traceability attack

We know that this type of attack can be performed if the fixed messages are transmitted between the parties, so the attacker can trace the message. In the proposed scheme, nonce values which are the random numbers generated by a statistically secure PRNG that is called "xorshiftR+" was used to prevent this type of attack. Sender and receiver or in other words both parties add these nonce values ($N_t$ and $N_a$) to all transmitted messages. Random numbers used as nonce values are important and must have some statistical properties so they can be used in this type of application for security concerns. xorshiftR+ is for this study because it is tested both in NIST STS and TestU01 and successful in both test suites (Çubuk et al., 2017). Adding $N_t$ and $N_a$ values to all transmitted messages before the encryption process both encrypted data and estimated MAC are also changed for each session. Therefore, we can define the proposed scheme as a secure one against traceability attacks.

### 4.2. Replay attack

A replay attack can be performed by recording the messages during the successful authentication process and sending them again to authenticate a non-valid party. It is also assumed that when the terminal issues a request, the attacker can substitute a valid party and participate in the charging or billing process instead of a car. In the proposed scheme, the attacker can resend the recorded messages to intervene in the process, but as can be seen in Figure 2, it is not possible to recreate a valid M3 value. M2 is created by using $N_a$ nonce value so it is always changing. This is because M3 values are encrypted by using M2 it is also continuously changing. Moreover, previously used Na values were saved on the server database so if the replay attack is tried server will respond to it as a replay attack easily.

### 4.3. Physical disclosure attack

In this type of attack, it is assumed that the attacker can physically gain access to vehicle, and all its hidden information. If we put it differently, it may be possible by stealing or physically seizing the vehicle. As soon as this situation is detected, the relevant vehicle registration will be disabled from the central server and there will be no charge for this vehicle. In addition, since the information of this vehicle will become completely unavailable, it will not be possible for another vehicle to initiate a charging process on a fake record with this information.

### 4.4. Impersonation attack

This attack is a type of fraud where the attacker poses as a trusted vehicle to manipulate the terminal and server, harm the system or capture sensitive information by violating confidentiality. In the proposed scheme, it is known that both M2 and M3 are encrypted. For encryption, the AES-256 algorithm is used. Considering the charging and billing process, the AES-256 algorithm is safe. With today's technical possibilities, it is not possible to break the AES-256 algorithm in the required time frame for these operations. On the other hand, the attacker cannot calculate the M2 value without having $ID_a$ and $k_a$. Moreover, both M2 and $k_g$ encryption key are required to calculate the M3 message. Also, it is not possible to calculate MAC value without having $k_a$ value. Because of all these reasons, the vehicle cannot impersonate the identity and it can be said that this scheme is secure against the impersonation attack.

### 4.5. Desynchronization attack

If there is a counter or a synchronized key between the parties and it is updated synchronously, this attack type can be applied. Also, it can be said that this attack can be caused by ensuring that the sequence number in the packets transmitted between the parties is different from the expected sequence number. If both parties have different counter values or keys so this occurs the synchronization problem. In this situation, transmitted packages can be discarded. Using this state, the attacker can send a packet with a correct sequence number and manipulate the system. Moreover, the attacker can direct or change the communication.

For this proposed scheme, there is no counter used on both the terminal and vehicle side or a key that needs to be synchronized for each session. Moreover, M3 and $N_a$ nonce values are taken by the terminal and M4 is created using $N_a$ and the values stored on the terminal side. $N_a$ random number which is used as nonce is also used to create M3. The attacker cannot change the M3 because it is encrypted using AES-256. MAC value guarantees the invariance of the M3 and other used values.

### 4.6. Denial of Service attack

A Denial of Service (DoS) attack is designed to render a serving device or network unusable. In this attack, the serving party is faced with demand beyond its capacity, causing the system to crash. After a successful attack, the serving system is temporarily or permanently unavailable. The presented scheme is not affected by Dos attacks. The main reason for this is that it is not particularly affected by the desynchronization attack and there is no structural problem on the service side that can be affected by this attack. Both the terminal and the vehicle are available to accept communication all the time.

### 4.7. Man in the Middle attack

Man in the Middle attack is possible if the data is publicly transferred between the parties. In the proposed scheme, the data is encrypted with AES-256. In addition, some additional security mechanisms and operations have been added to the scheme. Therefore, it can be said that the related scheme is safe against Man in the Middle attacks.

### 4.8. Cloning attack

Vehicle ID is not transmitted publicly between the vehicle and the terminal. This information is encrypted with AES-256. In addition, instead of the vehicle ID being encrypted alone, an XOR operation is made using a random number and then this result message is encrypted. This prevents the attacker from obtaining and cloning the vehicle ID information. AES-256 is an algorithm that is technically infeasible to break if the key is not compromised.

### 4.9. Eavesdropping

The attacker can listen to the communication between the vehicle and the terminal but the sent critical data is encrypted so the attacker can't estimate the plain text of that critical data.

## 5. CONCLUSION

Although the use of electrical energy in daily life has been common for many years, its use in vehicles has only become widespread in recent years. With this spread, vehicles such as automobiles, motorcycles, buses, pickup trucks and tractors, in addition to those working with fossil fuels, are also widespread in our daily lives. Considering the current battery technologies, these devices need to be charged frequently, even some of them every day. Charging on the street will become a necessity, especially in countries like ours, where individual and collective parking lots are insufficient. Systems should be recommended so that this charging process can be done safely, uninterrupted and anywhere. The lack of any standard on this subject has been the main motivation for the studies on this subject. In this study, a system that can provide a solution to all these problems and allow land vehicles to be charged on the street is proposed. In the system, an authentication scheme is proposed in order for the device to verify each other with the terminal and to ensure safe billing. A scheme proposal was also required for the authentication processes of electric vehicles, which eventually turned into a computer system. It is observed that the scheme successfully prevents well-known forms of attacks such as traceability, impersonation, replay, desynchronization, physical disclosure, man in the middle, denial of service, eavesdropping and cloning attacks.

In the future, this work can be implemented by governments and security tests can be carried out with a tool such as Scyther tool (Cremers, 2008; Cremers, 2014) or Tamarin prover (Meier et al., 2013) and the system can be improved with hardware improvements.